\documentclass[letterpaper,twocolumn,10pt]{article}
\usepackage{usenix2019_v3}

\usepackage{tikz}
\usepackage{amsmath}
\usepackage{tikz}
\usepackage{tabularx}
\usepackage{graphicx}
\usepackage{adjustbox}
\usepackage{comment}
\usepackage{tcolorbox}
\usepackage{multirow}
\usepackage{pifont}
\usepackage{subfig}
\usepackage[ruled,vlined]{algorithm2e}

\usepackage[autostyle=true]{csquotes} % 
\usepackage[maxcitenames=2]{biblatex}
\begin{document}
\date{}

\title{\Large \bf Being Single Has Benefits. Instance Poisoning to Deceive Malware Classifiers \\
}

\author{
{\rm Tzvika Shapira}\\
Ben-Gurion University of the Negev, Israel
\and
{\rm David Berend}\\
Nanyang Technological University, Singapore
\and
{\rm Ishai Rosenberg}\\
Ben-Gurion University of the Negev, Israel
\and
{\rm Yang Liu}\\
Nanyang Technological University, Singapore
\and
{\rm Asaf Shabtai}\\
Ben-Gurion University of the Negev, Israel
\and
{\rm Yuval Elovici}\\
Ben-Gurion University of the Negev, Israel
} 

\maketitle

\begin{abstract}

The performance of a machine learning-based malware classifier depends on the large and updated training set used to induce its model. 
In order to maintain an up-to-date training set, there is a need to continuously collect benign and malicious files from a wide range of sources, providing an exploitable target to attackers. 
In this study, we show how an attacker can launch a sophisticated and efficient poisoning attack targeting the dataset used to train a malware classifier. 
The attacker's ultimate goal is to ensure that the model induced by the poisoned dataset will be unable to detect the attacker's malware yet capable of detecting other malware. 
As opposed to other poisoning attacks in the malware detection domain, our attack does not focus on malware families but rather on specific malware instances that contain an implanted trigger, reducing the detection rate from 99.23\% to 0\% depending on the amount of poisoning. 
We evaluate our attack on the EMBER dataset with a state-of-the-art classifier and malware samples from VirusTotal for end-to-end validation of our work. 
We propose a comprehensive detection approach that could serve as a future sophisticated defense against this newly discovered severe threat.

\end{abstract}

\section{\label{introduction}Introduction} 

Both static and dynamic malware detectors have made great strides in the past few years with the emergence of machine learning (ML) and deep neural network (DNN) models~\cite{ucci2019survey}.
The ability of ML and particularly DNN models to learn complex concepts and patterns makes them highly effective in detecting even previously unknown malware~\cite{bekerman2015unknown,wang2019heterogeneous}.

While being very effective, in recent years ML and DNN models have shown to be vulnerable to adversarial learning attacks; these attacks have become a major threat and are being developed at a rapid rate. 
Adversarial ML attacks can be divided into two classes - evasion attacks and poisoning attacks.
In an evasion attack, the adversary attempts to create a carefully crafted adversarial sample that can fool the model~\cite{szegedy2013intriguing,carlini2017towards,goodfellow2014explaining}.
In a poisoning attack, it is assumed that the adversary has access to the model's training pipeline. 
Exploiting this access, the adversary poisons the dataset, and as a result, a model trained with these adversarial samples will make incorrect predictions on general or for specific samples~\cite{biggio2012poisoning,munoz2017towards,yang2017generative, suciu2018does}.

% Introduce concept of family poisoning
Poisoning attacks on malware detectors have been developed to launch denial-of-service attacks against the detector~\cite{chen2018automated,biggio2014poisoning} and to disable the detector in order to prevent it from detecting specific malware families (i.e., a malware family targeted attack).
In the latter case, the malware detector's detection accuracy remains almost unchanged, except in the case of the one poisoned malware family that the attacker wants to prevent from being detected.
This makes malware family targeted attacks difficult to discover when the malware detector tests all families together~\cite{suciu2018does,chen2018automated,sasaki2019embedding}.
At the same time, we claim that this is an inherent drawback of the malware family targeted attack, as once checked by all families individually, such poisoning attacks are easy to detect.
Testing all families can be accomplished due to the relatively small number of instances required (a few for each family) to periodically validate that the detector's newly induced model can successfully detect each malware family.

% BEGINNING OF PICTURE POISONING
An attack that is much more difficult to detect involves poisoning a specific instance of a malware family, rather than poisoning the entire family. 
Such attacks are a recent development in the malware domain. To the best of our knowledge, \citeauthor{severi2020exploring} were the first to introduce the instance targeted attack to the malware domain~\cite{severi2020exploring}; while successful on limited data of the benchmark set, the authors craft their attack assuming an adjustable feature space, making the attack irrelevant for many real-world end-to-end cases.
In the computer vision (CV) domain, however, state-of-the-art instance targeted poisoning attacks with real-world applications have been extensively studied~\cite{shafahi2018poison, liu2017neural, liu2017trojaning}; such attacks have been shown to be successful while maintaining the ability to classify the rest of the instances.
These attacks are recently extended (within the CV-domain) and demonstrate the ability to create an instance targeted approach using a backdoor trigger, which causes the model to misclassify just a specific instance that contains the backdoor trigger, while the rest of the instances are classified normally~\cite{gu2017badnets,liu2017trojaning,chen2017targeted}. 
Such attacks are very difficult to detect, due to both the natural behavior of the model on all instances except one, and the fact that the defender is unable to realize that he/she is under attack when training the detection model.

Inspired by the CV-domain, this paper introduces the above mentioned concepts to the malware domain and aims to create a novel instance targeted poisoning attack which is capable of targeting only malware instances with a specially crafted trigger; we demonstrate that these malware instances \textit{are not} detected by the malware detector in the testing phase, whereas the rest of the malware family members \textit{are} detected, which gives the defender false confidence in his/her detector. 
Our attack operates by choosing the benign samples that are most similar to the target malware instance and adding features to the benign and targeted malware samples (which act as the backdoor trigger) that do not affect the samples' functionality. 
The model is re-trained with the modified benign samples.
Then, during the testing phase, when introduced with the malware sample that carries the backdoor trigger, the targeted instance is classified as a benign sample, while the other samples of the malware instance's family are still classified correctly as malware as they do not contain the trigger features. 
Our results confirm the effectiveness of the attack by showing that the malware instance attack is 100\% undetected when the poisoned benign samples added represent less than 4\% of the entire training set in the best case scenario and 7.5\% on average.

We claim that this attack is feasible, since security companies invest a significant amount of effort and resources in staying up-to-date on the latest threats, by retraining their malware detector with new files from various public sources (such as VirusTotal~\cite{VirusTotal}). 
This serves as an exploitable target that can be used to add poisoned benign files with a backdoor to the public sources, thereby poisoning the majority of state-of-the-art malware detectors.
Moreover, after training the detector with the poisoned dataset, the defender may not even be aware that the detector has been compromised as most of the detector's classifications remain intact and will not behave in a suspicious manner.

We evaluate our novel malware instance poisoning attack on the EMBER benchmark dataset~\cite{anderson2018open} and validate its effectiveness on real-world malware samples provided by VirusTotal, which is the main public source of malware samples used by security companies to create their malware detector training set~\cite{virustotal_source}.

The identified threat calls for an urgent solution; to contribute to this, we provide a comprehensive analysis of best practices from the malware, CV and deep learning (DL) domain. For the malware domain we examine the application by a novel family-based detection technique which can detect up to 99.23\% of malware family targeted poisoning attacks, while \emph{underperforming} in detecting instance targeted attacks, which gives attackers a good reason to resort to instance targeted attacks as an attack angle. Based on the the CV-domain where the instance targeted attack originated, we propose utilizing an autoencoder-based approach, which we train on the clean benign files to detect the poisoned malware using the reconstruction error.
Finally, for the DL domain, we introduce out-of-distribution (OOD) detection to the malware detection sphere, which leverages the neural activation behavior to detect poisoned malware instances.
While not yet fully developed, the proposed instance-based detection methods can serve as the basis of a defense mechanism for which the OOD-detection seems the most promising of mitigating the instance poisoning attack we revealed in this study, an attack which state-of-the-art detection systems are currently unable to detect.

To summarize, the contributions of this paper are as follows:
\begin{itemize}
  
    \item We introduce a novel malware instance poisoning attack, inspired by the CV-domain, which has a 100\% success rate of evading detection by current malware detection models poisoning as little as 7.6\% of the training set. 
    
    \item We demonstrate a successful end-to-end evaluation on the EMBER dataset and on real-world malware samples obtained from VirusTotal for threat model validation, which can impact all current malware detector companies.
    
    \item We provide a comprehensive review of best practices for defending against poisoning attacks for both family-based and instance-based scenarios, and we present a simple yet effective novel detection technique for family-based attacks that achieves 99.23\% accuracy. 
    However, this detection technique fails for the instance targeted attack, which remains a challenge that is currently unaddressed by state-of-the-art detection methods.
    
    \item We address the instance-based  detection challenge by examining best practices from malware, CV and DL domains, further introducing autoencoders and out-of-distribution detection, to give a comprehensive view on the most promising directions to mitigate this new threat.
    
\end{itemize}

Our novel malware instance poisoning attack presents a serious new threat to the security field for which we provide a comprehensive review and promising direction to address our novel attack.
Our large-scale evaluation of both the attack and the defense angles underscores the need to increase awareness regarding this threat, and moreover, the urgency of developing defense mechanisms against it.

\section{Related Work} \label{relate_work}

\subsection{Poisoning Attacks}

While ML and DNNs are considered highly effective in solving many tasks, it has been observed that these models have an inherent weakness and lack robustness towards adversarial attacks. \citeauthor{szegedy2013intriguing}~\cite{szegedy2013intriguing} were the first to show that deep neural networks are highly susceptible to adversarial attacks.
Inspired by their work, further research demonstrating different attack and defense methods followed~\cite{goodfellow2014explaining,kurakin2016adversarial,papernot2016limitations,papernot2016distillation}. 
While most of the attacks focused on the CV-domain, some of the studies presented attacks targeting malware detection models~\cite{rosenberg2018generic, grosse2017adversarial, hu2017generating, kreuk2018adversarial, kolosnjaji2018adversarial}.
Most of the research focused around evasion attacks, which do not modify the model at all but rather modify the adversarial example at inference time, exploiting the model's behavior and attributes.

As the research around evasion attacks increased, some researchers also focused on poisoning attacks~\cite{biggio2012poisoning, munoz2017towards, yang2017generative}. 
Poisoning attacks focus on both the training and inference phases of the model's pipeline and attempt to change the model itself during the training phase, in order to deceive it in the inference phase. 
\citeauthor{biggio2012poisoning} used poisoning to attack support vector machines, by computing gradient ascent based on the properties of the SVM's optimal solution~\cite{biggio2012poisoning}. 
Attacks against neural networks were proposed several years later, showing that poisoning attacks can have devastating results in these cases. \citeauthor{munoz2017towards} proposed a poisoning attack that uses back-gradient optimization and proved that this attack can influence a broad range of models that use gradient-based learning operations, including neural networks~\cite{munoz2017towards}. 
In~\cite{yang2017generative}, the poisoning data was also generated using gradient-based methods, and the data generation was accelerated using a GAN~\cite{goodfellow2014generative}.

Initial poisoning attacks against neural networks focused on harming the neural network's functionality and demonstrated their failure when poisoned against larger groups of malware or targeted malware families. 
Although rarely addressed in the malware domain, instance poisoning has recently gained a great deal of attention in the CV-domain; in this case, the attacker does not want to cause the neural network to fail completely -- just specific, targeted cases, in order to prevent the attack from being detected. 
\citeauthor{shafahi2018poison} crafted a single class attack against CV models, focusing on transfer learning \cite{shafahi2018poison}. 
Their attack is targeted and does not require the attacker to control data labeling. 
The poisoned instances are created by identifying a point that is both close to the target class in the feature space and close to the source class in the input space. 
Their attack is highly effective in the scenario of transfer learning, requiring only a single poisoned instance for 100\% attack success; In the end-to-end training scenario, the attack requires an increased number of poisoning examples, in addition to watermarking the poisoned instances.
In~\cite{chen2018automated}, the poisoned examples are crafted using a Jacobian matrix~\cite{papernot2016limitations} and are used to poison different Android malware detectors, reducing their accuracy dramatically on the targeted type of malware.
\citeauthor{suciu2018does} created a sophisticated cross-platform attack that targets a specific malicious Android application and used it to attack an SVM-based model~\cite{suciu2018does}. 
Their attack affected the specific malware application and other similar applications. 
The authors demonstrated their attack in other domains and evaluated it using the FAIL attacker model, which characterizes the attacker's knowledge.

Another type of poisoning attack from the CV-domain uses a backdoor, or a Trojan trigger, to misclassify specific poisoned instances. 
few studies attempted to convert the backdoor attacks to the malware domain, focusing on family poisoning attacks. 
In~\cite{sasaki2019embedding}, the converted backdoor attack is used as a family targeted attack, which utilizes the same optimization problem used in~\cite{munoz2017towards}. 
Their attack does not work on all malware families because of the varying difficulty of creating the poisoned instances.
Furthermore, their attack is only compatible with feature-vectors, requires that a rather large proportion of the dataset is poisoned, and was not evaluated in an end-to-end scenario. 
\citeauthor{severi2020exploring} presented a backdoor attack on malware families that uses explainable ML techniques~\cite{severi2020exploring} and evaluated their attack using the EMBER dataset~\cite{anderson2018open}.
Their attack achieved excellent success rates with minimal injections of poisoning samples, but this is only the case when it is a pure feature vector attack. 
When tested with real-world constraints, their attack is far less effective as we carefully examine in a comprehensive comparison of the work of \citeauthor{severi2020exploring} to our work in Section~\ref{previous_work_comparison}.

\subsection{Defense Mechanisms}

To the best of our knowledge, detection methods for poisoning attacks in the malware domain have not been addressed for family-based or instance-based attacks. 

To address our proposed instance-based attack comprehensively, we must look to the CV-domain where various detection methods against instance-based backdoor poisoning attacks have been suggested.
\citeauthor{liu2018fine} used "fine-pruning," a combination of pruning and fine-tuning, to create a model robust against backdoor attacks~\cite{liu2018fine}. 
In~\cite{liu2017neural}, the authors presented three different defense methods: input anomaly detection, retraining, and input preprocessing. 
These initial methods achieved reasonable results, but they have a few shortcomings, such as a reduction in the original model's performance, and high complexity and computation costs.

In~\cite{wang2019neural}, the authors proposed \textit{Neural Cleanse}, which detects a backdoor by locating an outlier among backdoor candidates for all of the different classes, identifying the target of the backdoor and blocking or removing it. \citeauthor{gao2019strip} introduced STRIP \cite{gao2019strip}, a method that uses intentional perturbations on the inputs in order to recognize the randomness of the prediction and to detect backdoors when the predictions are not random. 
In~\cite{chen2019deepinspect}, the backdoor is detected by measuring the perturbation distance required to shift the target class. 
When a backdoor is present, the distance of the perturbation to its target class is significantly smaller than other perturbation distances. 
In~\cite{liu2019abs}, the backdoor is detected using neuron behavior analysis. 
The neurons of the network are stimulated, enabling the neurons that elevate the output of a specific class to be located, thus revealing a backdoor. These defense methods rely on domain specific characteristics and therefore only apply to the CV-domain.

To summarize, previous attempts for detecting instance-based backdoor attacks are either irrelevant for the malware domain or have significant computational overhead with inadequate results, which calls for the development of compatible detection methods capable of addressing the novel instance-based attack.

\section{Threat Model} \label{threat_model}
\begin{figure}
    \centering
    \includegraphics[width=\columnwidth] {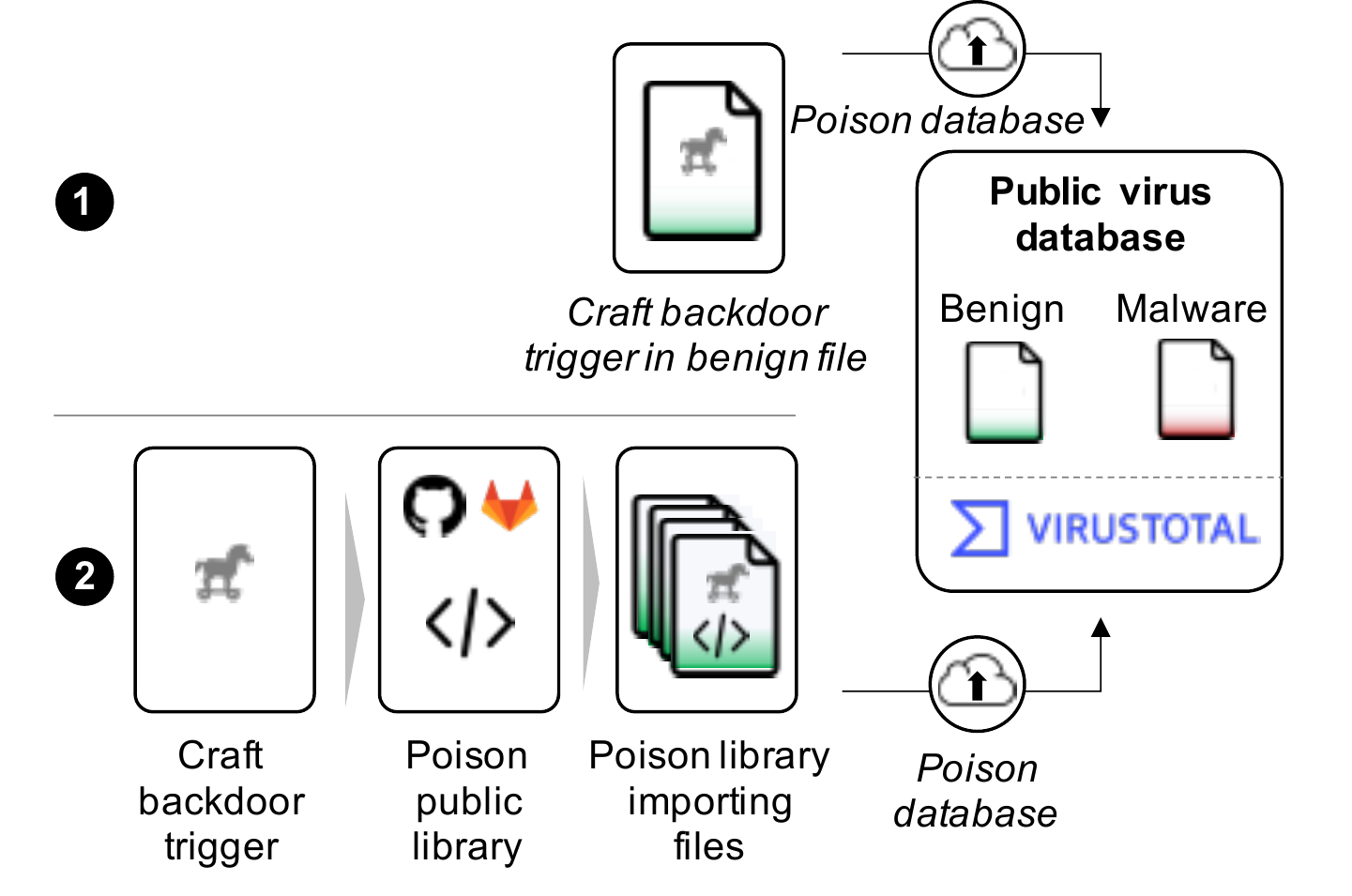}
    \caption{Overview of threat model.} 
    \label{fig:threat_model}
\end{figure}

Security companies invest significant means and efforts towards being up-to-date with the latest threats, for which they retrain their malware detector training set with new files from various public sources (such as VirusTotal~\cite{VirusTotal}). 
Since these sources are public, they allow third-parties, e.g., the attacker to add benign files, which incorporate the backdoor for poisoning the targeted malware-instance. 
The more benign files with the backdoor trigger are uploaded to the public source, the better the success rate of the attack to remain undetected by the poisoned malware detector (path \ding{182}, Figure~\ref{fig:threat_model}).

While the attacker may be able to upload benign files with the backdoor himself in path \ding{182}, this approach is limited in its capacity, as a few million of files are required for a 100\% successful attack. 
However, it can be accomplished through large botnets with IP masking efforts and the backdoor trigger feature-integration in various benign files which have a different functionality to remain undetected by low-concentration of specific benign-file types. 
Thereby, public sources, such as VirusTotal will not be able to detect the additional benign files as such databases experience millions of uploads daily with different benign and malware files~\cite{virustotal_source}. 
Hence, we further propose a second threat model for scale, where the complexity of integrating the backdoor in the benign files is higher, but the amount at which different poisoned benign files from various sources can be uploaded is also increased. Following path \ding{183} in Figure~\ref{fig:threat_model}, the attacker makes a contribution to a commonly used public code library while hiding the backdoor in his contributed code, which imposes additional complexity. 
Optimally, this contribution addresses a newly discovered vulnerability and proposes a fix, while hiding the backdoor inside the vulnerability-fix, which is updated to all library-incorporating benign files. 
This results in the public code library being used by different software products, where some of them will be uploaded by third-parties to public virus databases. 

Both methods demonstrate the effectiveness of the attack, where large companies or nation-states have the ability to poison public databases used by security companies, while remaining undetected in both the poisoning effort and the later instance attack itself.

\section{Methodology}\label{sec:proposed_method}

In this section we present our novel malware instance poisoning attack and explain the necessary concepts and algorithms in context of the threat model. 
We also explore three potential detection mechanisms from malware, CV and DL domain to evaluate the overall implications of our attack.

\subsection{Novel Malware Instance Poisoning Attack} \label{sec:attack}

\begin{figure*}
    \centering
    \includegraphics[width=1.0\textwidth] {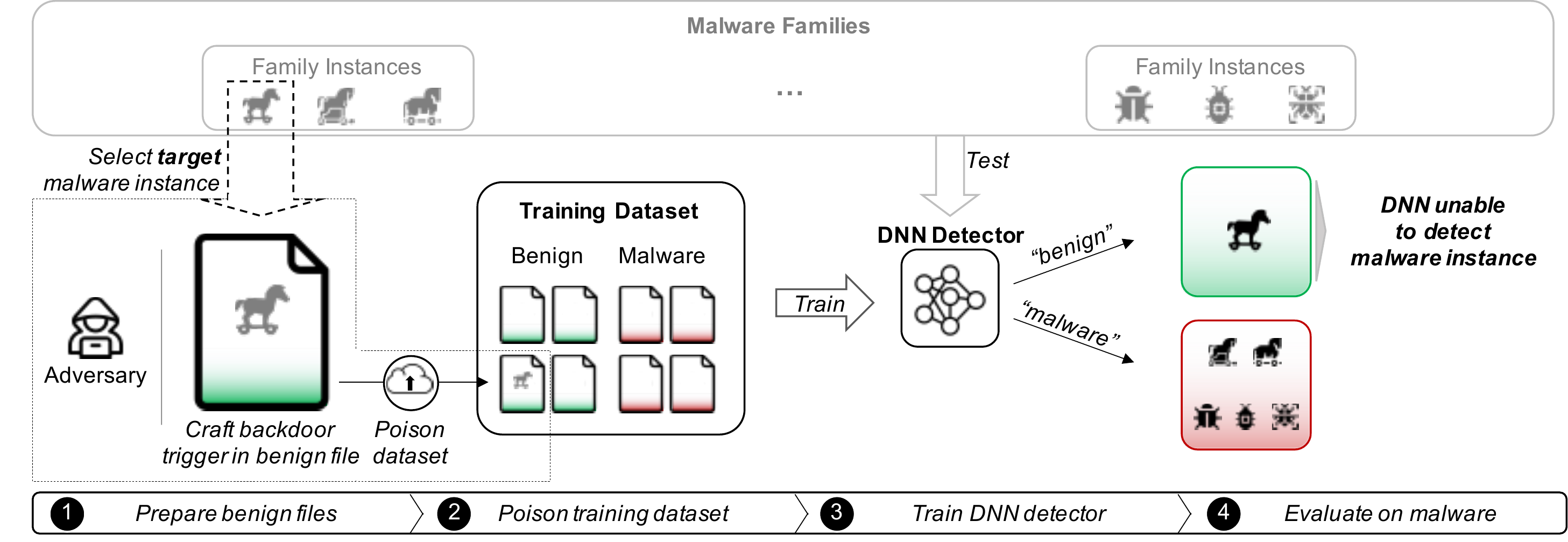}
    \caption{Overview of our novel instance-based attack.} 
    \label{fig:attack_framework}
\end{figure*}

Since the majority of recent poisoning attacks against malware classifiers are family targeted attacks, we first present a family targeted attack based on the poison frog attack~\cite{shafahi2018poison}, which originates in the CV-domain and presents the baseline comparison for our instance attack.
For the family attack, for each malware sample in the target family, the attack locates the closest benign example in the feature space using the Euclidean distance and creates a poisoned example by moving toward the benign example (using the gradient), while limiting the change in the input space. The attack attempts to optimize the following problem:
\[\mathop{argmin}_{x}\|f(x) - f(t)\|_2^2 + \beta\|x - b\|_2^2\]

The right side of the equation is the distance between the poisoned sample $x$ and a benign sample $b$ in the input space (limiting the change in the input space), and the left side of the equation is the distance between the poisoned sample $x$ and the target malware sample $t$ in the feature space. The procedure for solving the optimization problem uses a forward-backward splitting iterative process~\cite{goldstein2014field}, which utilizes gradient descent to minimize the distance in the feature space, and a proximal update to minimize the distance in the input space.
Eventually, these poisoned examples are used as part of the process of retraining the model in order to fool the model during the inference phase.

This family targeted attack is a white-box feature vector attack, without restrictions on the features that the attacker chooses to change. Since some of the features in a PE file cannot be modified without harming the malware's functionality (e.g., bytes in the malware's code section), not all of the malware generated in this way would be able to run properly.

Inspired by instance-based attacks originating in the CV-domain, such as the BadNet attack~\cite{gu2017badnets} and the Trojaning attack~\cite{liu2017trojaning}, we propose a novel malware instance poisoning attack (Figure \ref{fig:attack_framework}).
Our attack differs from most existing poisoning attacks in the malware domain by targeting a single instance of malware, instead of a whole malware family.

Using our attack as illustrated in Figure~\ref{fig:attack_framework}, the attacker can bypass the defender without being detected while maintaining the accuracy for the rest of the instances, even for those other instances within the poisoned instance's malware family. This may allow our novel attack to bypass state-of-the-art detection mechanisms which are trained on detecting the family-based poisoning scenario. In detail, the main actions for the adversary take place in the first two steps. In \ding{182}, the attacker selects a malware instance he wants to target and crafts the backdoor trigger in the benign files which will allow him later on to bypass the detection technique for the selected malware instance. After crafting the benign samples, step \ding{183} illustrates the upload of the crafted benign files to a public training dataset, which are commonly used by defenders to train their model (see Section \ref{threat_model}, Threat Model). Step \ding{184} illustrates the defender training the DNN model using the poisoned dataset without being aware that his/her model is poisoned to fail for the targeted malware instance. Finally, step \ding{185} illustrates the desired outcome of the attack, in which the targeted malware instance is now predicted by the defender's model as benign file while all other instances are predicted as malware, demonstrating the success of the attack. 

For a successful integration of the CV originating instance-based attack into the malware domain, differences towards classification type, functionality, and feature types must be addressed first:

\begin{enumerate}
    \item\textbf{Classification type}. In contrast to the CV-domain, which has many different target class possibilities, in the malware domain a binary classification problem is present. This forces us to use only the benign class for the poisoning process rather than a variety of target classes.
    \item\textbf{Malicious functionality}. In order to deceive a malware detector, the malware instance itself needs to be altered with a small perturbation, while maintaining the original functionality of the executable/malware. On the other hand, in the CV-domain, pixels of an image can be changed in any desired degree without harming the 'functionality' (semantic validity) of the image. Hence, in the malware domain the perturbation location needs to be carefully chosen, so that functionality of the target instance is maintained.
    
    \item\textbf{Feature types}. In the CV-domain, an image is a fixed matrix of pixels which all have the same properties with different numerical values. Malware detectors use various types of features with different properties (e.g., discrete features), which vary in length and size and are difficult to perturb. In addition to choosing a location that maintains the functionality, the way in which the backdoor for the attack is integrated also varies by location. 
\end{enumerate}

To address the feature types challenge of crafting the backdoor trigger into a benign file while remaining functionality of the file we present Algorithm \ref{algAttack}. A set of benign files $B$ from the dataset $D$ is taken (the relative size of the poisoned set will be discussed later in Section \ref{sec:novel_attack}) and modified by adding, e.g., sections to a PE file  that do not have any influence on the functionality of the file but would act as the trigger for our poisoning attack, thus creating the poisoned set $B^*$. We also modify our target malware $t$, which we want to "hide from the model," and add the same sections to its PE file, creating $t^*$. $t^*$'s functionality is not affected in the process. By choosing to add sections that do not affect the functionality to the files, we address both the malicious functionality challenge and the feature type challenge. 

In order to create an efficient poisoned set, for the benign files for $B^*$ we choose the benign files closest to the target malicious instance $t$. 
The benign files are chosen for $B^*$ in order, starting with the file that has the smallest Euclidean distance from $t^*$ (this is relevant for the case of the first attacker presented in Section~\ref{threat_model}). 
We also attempt to randomly select the benign files for poisoning (as in the case of the second attacker presented in Section~\ref{threat_model}), but the performance of the attack decreases, as will be discussed in Section \ref{sec:novel_attack}.

The model $M$ is then retrained with the poisoned set $B^*$, and the new poisoned model $M^*$ is created. During the inference phase, the accuracy of $M^*$ stays the same for the full dataset $X$, but changes for $t^*$:
\[M(X-t) = M^*(X-t)\] 
\[M(t) \neq M^*(t^*)\]

The main drawback of this method is the utilization of only a narrow fraction of the features for the poisoning process. As opposed to the CV-domain, where there is a large variety of backdoor triggers available in the feature space, in the malware classification domain we must handpick the poisoning features carefully in order to create a physically feasible attack that maintains all of the malware functionality. In order to make the data poisoning effective with a small portion of features, a larger number of files needs to be used in the poisoning process itself, and the ratio of the poisoned data $B^*$ from $X^*$ is significantly larger. 

\begin{algorithm}
\DontPrintSemicolon
    \KwIn{$M$: the model,\newline
    $D$: the dataset (benign files),\newline
    $r$: the relative size of the poisoning set, \newline
    $s$: poison section detail,\newline
    $k$: the neuron enhancement factor (optional), \newline
    $top$: the number of neurons to enhance, \newline
    $t$: the target malware}
    \KwOut{$M^*$: the poisoned model,\newline
    $t^*$: the new target malware}
    \tcp{Construct poisoning set according to distances}
    $M_D \gets $ \textsc{SortedDistances($D, t$)} \\
    $B \gets \{\}$ \\
    \While{$|B| < (|D| * r)$ }{
        $B.add(M_D.next()$
    }
    \tcp{Add poisoning sections (trojan trigger) to poisoning set and target malware sample}
    $B^* \gets $ \textsc{AddPoisonSections($B, s$)} \\
    $t^* \gets $ \textsc{AddPoisonSections($t, s$)} \\
    \tcp{Training the model on poisonrd data}
    $M^* \gets $ \textsc{TrainModel($M, B^*$)} \\
    
    \KwRet{$M^*, t^*$}
    
\caption{Novel malware instance poisoning attack}\label{algAttack}
\end{algorithm}

\subsection{Proposed Defense Mechanisms}

In this work we present a novel malware instance poisoning attack for which no detection method has been proposed. Since malware family poisoning attacks are addressed in related studies, we start by presenting a detection method following existing concepts against these attacks to identify different detection options that may be effective for the instance-based scenario. These are an autoencoder-based detection technique originating from the CV-domain and an out-of-distribution detection technique originating from the DL-domain. Thereby, we approach the attack from all three functional perspectives: Malware for direct domain, CV for origin and DL for underlying nature of the model.

\subsubsection{Detecting Family Targeted Attacks}\label{sec:meth:family_detect}

Previous poisoning attacks in the malware detection domain have been shown to be effective in fooling malware detectors for specific malware families, while maintaining a high detection rate for other malware families~\cite{suciu2018does, chen2018automated, sasaki2019embedding}.
Related studies from the CV-domain proposed detection techniques that either do not apply to the malware domain, or have heavy computational overheads, including \cite{liu2017neural}, which presented a technique that requires a special  model to test all inputs. To address such drawbacks, it makes sense to construct a lightweight and malware-domain applicable detection method for family poisoning detection. 

We propose a family-based defense mechanism that is capable of detecting global ("across the board") poisoning attempts  as well as family targeted poisoning attacks against malware detectors, while maintaining performance and a steady runtime.

\begin{figure}
    \centering
    \includegraphics[width=1.0\columnwidth] {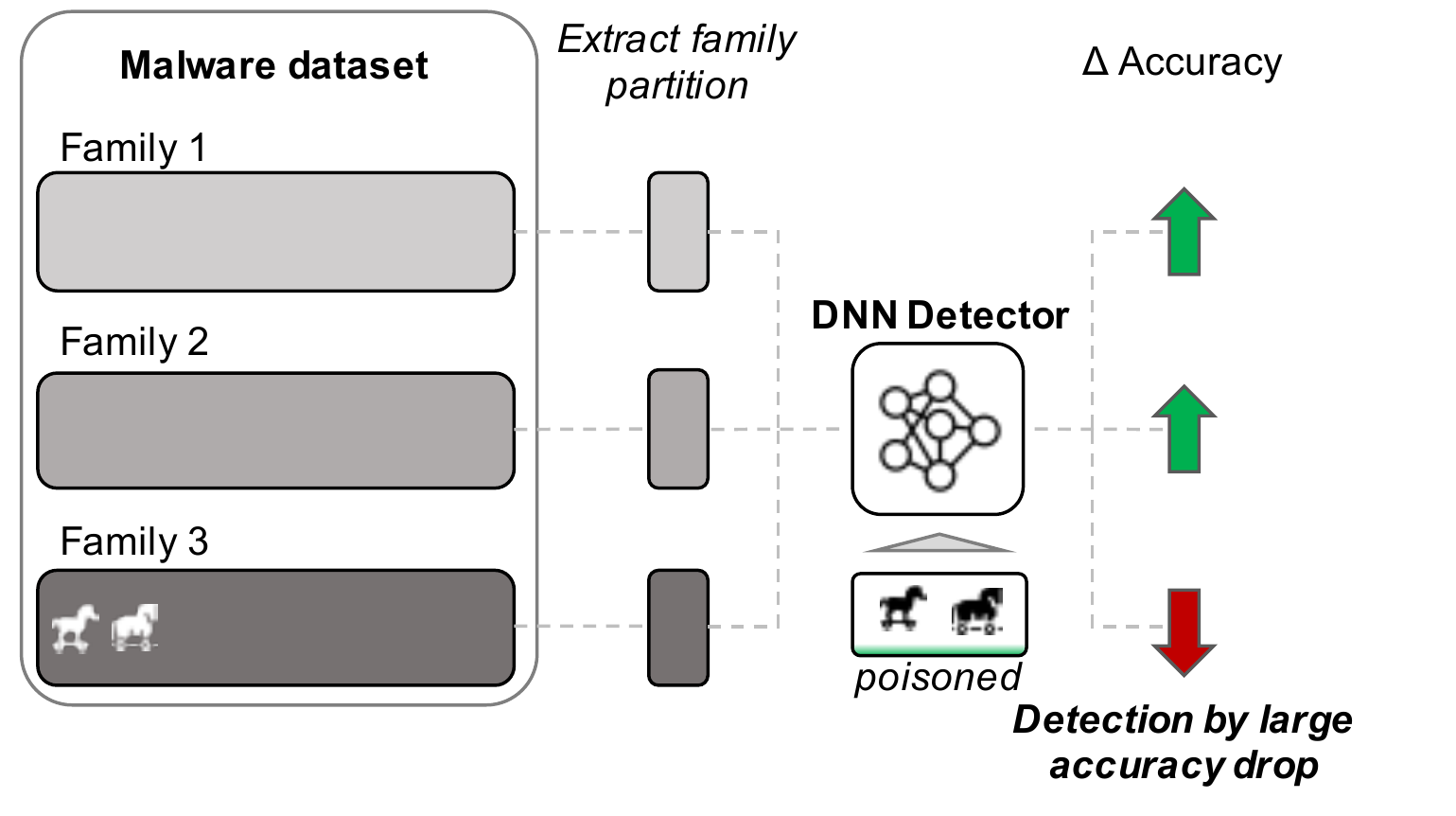}
    \caption{Overview of our proposed family-based detector.}
    \label{fig:defense_framework}
\end{figure}

Our technique uses a small representative subset of each malware family to create a verification set for the model. The technique is described in Figure \ref{fig:defense_framework}. For each malware family $ X_i \in \{ X_1 ... X_F\}$, we randomly choose $n$ examples to create a small partition $E$ of size $n*F$. The partition is tested on the detection model. The scores for each subset of a malware family $\{S_1...S_F\}$ are stored for future examination. The pseudocode used to craft the partition is shown in Algorithm \ref{algExamination}. After each of the detection model's retraining cycles, the partition is newly extracted, tested on the detection model, and new scores $\{S_1'...S_F'\}$  are collected. The new scores are compared to the original scores. If one (or more) of the malware family's current scores $S_i'$ drops more than the defined threshold $D$, it is an indication of a family targeted poisoning attack. Hence, an indication will be raised only when $S_i - S_i' > D$. The pseudocode of the detection is shown in Algorithm \ref{algDetector}. 

The proposed technique is simply crafted and only needs to be assembled once. The sizes of $n$ and $D$ are also determined once, and used repeatedly for detection. The size of the partition itself can be small and does not depend on the size of the whole dataset but rather on the number of different malware families there are to be tested. Real-world malware datasets have a significantly smaller number of different families than malware samples - for a dataset of size $N$, $F \ll N$, which is also true for the size of the partition: $F*n \ll N$. Our technique exploits this fact and uses it to create a lightweight, fast, and reliable family-based detector.

\begin{algorithm}
\DontPrintSemicolon
    \KwIn{$X$: full malware dataset,\newline
    $M$: the trained model,\newline
    $n$: the number of samples per partition.}
    \KwOut{A verification set $E$ and its initial score dictionary $S$}
    $S \gets \{\}$ \\
    \tcp{Randomly select $n$ samples for each malware family}
    $E \gets$ \textsc{RandomFamilyPartition($X, n$)} \\
    \ForEach {$X_f$ in $E$} {
        \tcp{Get the model predictions for the family $f$}
        $S_f \gets M(X_f)$\\
        \tcp{Store the predictions in the dictionary}
        $S[f] \gets S_f$\\ 
    }
    \KwRet{$E, S$}
\caption{Crafting the verification-set for family-based detection}\label{algExamination}
\end{algorithm}

\begin{algorithm}
\DontPrintSemicolon
    \KwIn{$M^*$: the model, retrained with new data,\newline
    $E$: the verification set, \newline
    $S$: the verification set scores, \newline
    $D$: threshold for accuracy drop}
    \KwOut{A boolean, indicating whether the model was poisoned or not.}
    \ForEach {$X_f$ in $E$} {
        \tcp{Get the predictions for each family with the newly trained model}
        $S_f' \gets M^*(X_f)$\\
        \tcp{Check if the score decreased significantly for each family}
        \If{$S[f] - S_f' > D$} {
            \KwRet{$True$}
        }
    }
    \KwRet{$False$}
\caption{Family-based detection technique}\label{algDetector}
\end{algorithm}

\subsubsection{Detecting Instance Targeted Attacks}\label{sec:instance_detect}

In the previous subsection we proposed a family-based detection technique to identify poisoned malware families. In this section, we explore how the family-based practices can be applied to the instance-based scenario. In addition, we provide insight into detection techniques originated in the CV-domain and introduce DL-based out-of-distribution detection to the malware domain to prove a comprehensive overview of promising directions for addressing the threat at hand.

\textbf{Family to instance detection.} In our initial attempt at detecting instance-based poisoning attacks, we follow the same methodology presented above for family-based attacks. Therefore, each time the model is retrained, we evaluate $\Delta accuracy$ on the family partition to detect if the result caused by the poisoned malware instance exceeds threshold $D$. A preliminary aspect for this scenario is that the instance-based attack needs to affect different members of a malware family for this detection method to be sufficient.

\textbf{Computer vision to malware detection.} Next, we focus on existing techniques in the CV-domain. Of these, utilizing the autoencoder reconstruction error seems the most promising~\cite{liu2017neural} as outlined in Section \ref{relate_work}.
In this method, the autoencoder learns the dataset's main features and reflects them in the low-dimensional latent space. Afterward, the input is reconstructed based on the main activated features from the clean benign files. In the case of examining a poisoned malware instance, a high reconstruction error is likely to be observed, as the poisoning backdoor trigger may be unable of being reconstructed, because it was not learned as a main feature of the clean benign dataset.

\textbf{Out-of-distribution detection.} The third and final approach for detecting instance-based poisoning attacks is out-of-distribution (OOD) detection; the baseline of this approach was originally proposed by Hendrycks et al.~\cite{ood:baseline} for distinguishing between two different datasets. Out-of-distribution is defined by dataset $A$, which has a data distribution $D_{IN}$ and another dataset $B$, which has a data distribution $D_{OUT}$. Assuming a DNN model is trained on $A$ 
(which in this case consists of clean benign files), it will be less likely to handle inputs from $B$ (in this case the poisoned malware instance) correctly if the corresponding distributions $D_{IN}$ and $D_{OUT}$ are different . For both datasets, OOD scores are extracted, forming the corresponding distribution. If the distribution of $B$ does not follow the distribution of $A$, then $B$ is considered out-of-distribution and vice versa, which means poisoned malware instances may be detectable. 

To calculate the OOD-score, we follow the baseline algorithm~\cite{ood:baseline} where the OOD score $s$ is calculated using the subtraction of 1 and the maximum softmax probability of the vectorized DNN output $\textbf{x}$  of length $K$ as  

\[    s(\textbf{x}) = 1 - argmax \left (  \frac{e^{x_i}}{\sum_{j=1}^{K}e^{x_j}}  \right )\] 

Finally, we extract the OOD-score from the clean benign samples to receive $D_{IN}$ and determine a threshold $N$ corresponding to the $N$-percentile of $D_{IN}$. If an OOD-score from $D_{OUT}$ is exceeding the value at the $N$-percentile of $D_{IN}$, we classify it as OOD and thereby as poisoned malware instance.

\section{Evaluation} \label{evaluation}

Our evaluation follows the methodology presented in the previous section, by first evaluating the novel instance attack and providing insights into its effectiveness and implications. Then, we move on to the defense perspective and examine the family-detection technique which is used as the foundation to address the instance-based scenario. For the instance-based detection, we evaluate the three proposed techniques to give a comprehensive understanding on addressing this new threat.

\subsection{Experimental Setup}

\subsubsection{Datasets}

The evaluation of our proposed attack and detection technique was performed using the most recent and comprehensive EMBER benchmark dataset~\cite{anderson2018open}. The EMBER dataset is a benchmark dataset of static PE file features.
The dataset contains 800K training samples, of which 200K are unlabeled. We omit the unlabeled samples resulting in 600K training samples and an additional 200K test samples. 
Half of the samples in the training and test sets are malicious, while the other half are benign. 
Each malware example contains 2,381 static features, extracted using the LIEF library~\cite{lieflibrary}. The features are a combination of byte histograms and PE structural metadata (e.g., sections, header information, imports, and exports), which are vectorized and normalized.

In addition, we reached out to VirusTotal, one of the main public virus database repositories, which is used in most detection software. VirusTotal provided real-world malware instances, and 20 samples are utilized in this work to validate the impact and effectiveness of our novel instance targeted attack in an end-to-end setting. 

\subsubsection{Malware Detection Models}

The experiments are conducted on two different models for comparison reasons:

\emph{The LightGBM model}, introduced by the EMBER project~\cite{anderson2018open}, achieves an accuracy score of 97.87\% and AUC score of 97.21\% on the EMBER dataset and serves as baseline. 

\emph{A static DNN model} obtains state-of-the-art results on the EMBER dataset and reflects modern day practices of malware detection systems. Following related work~\cite{rosenberg2020generating}, we use a feedforward neural network that contains two dense layers, each with 128 neurons and rectified linear activation units (ReLU)~\cite{relu} as activation functions. After each layer, we apply dropout of 0.2~\cite{relu}. 
This model achieves 95.72\% accuracy and an AUC score of 95.64\% on the EMBER dataset. 

\subsection{Attack Evaluation}

To compare the novel malware instance poisoning attack, we first conduct an evaluation of the family-based attack presented in Section \ref{sec:attack} which serves as a baseline for comparison to the instance-based setting, both in attack and defense scenario.
We then evaluate the novel instance-based attack. The attacks presented are black-box attacks, which work independently of the target models and thereby further increase potential impact, as they are able to target a wide variety of detection systems.

\subsubsection{Family-Based Attack}

Our goal in evaluating the family-based attack is to understand the strengths and limitations of the state-of-the-art malware poisoning practices, which we evaluate on the EMBER dataset for two different scenarios:

\emph{Known malware family.} In this setting, benign files are poisoned towards a malware family which is part of the training data and therefore known by the model. Despite the fact that the model is initially trained to detect the specific malware family, the attack attempts to fool the model using the poisoning such that it is unable to detect this family's malware samples after training.

\emph{Unknown malware family.} In this setting, benign files are poisoned towards a malware family that the model is not exposed in the training stage, meaning that the targeted malware family is not part of the training data. In this case, the model does not learn the features of the target malware family and aims to detect it based on features learned from relatively close malware families instead. In this scenario, the model is more vulnerable to a poisoning attack, as the family is not known to the model.

\begin{figure}
    \centering
    \includegraphics[width=\columnwidth] {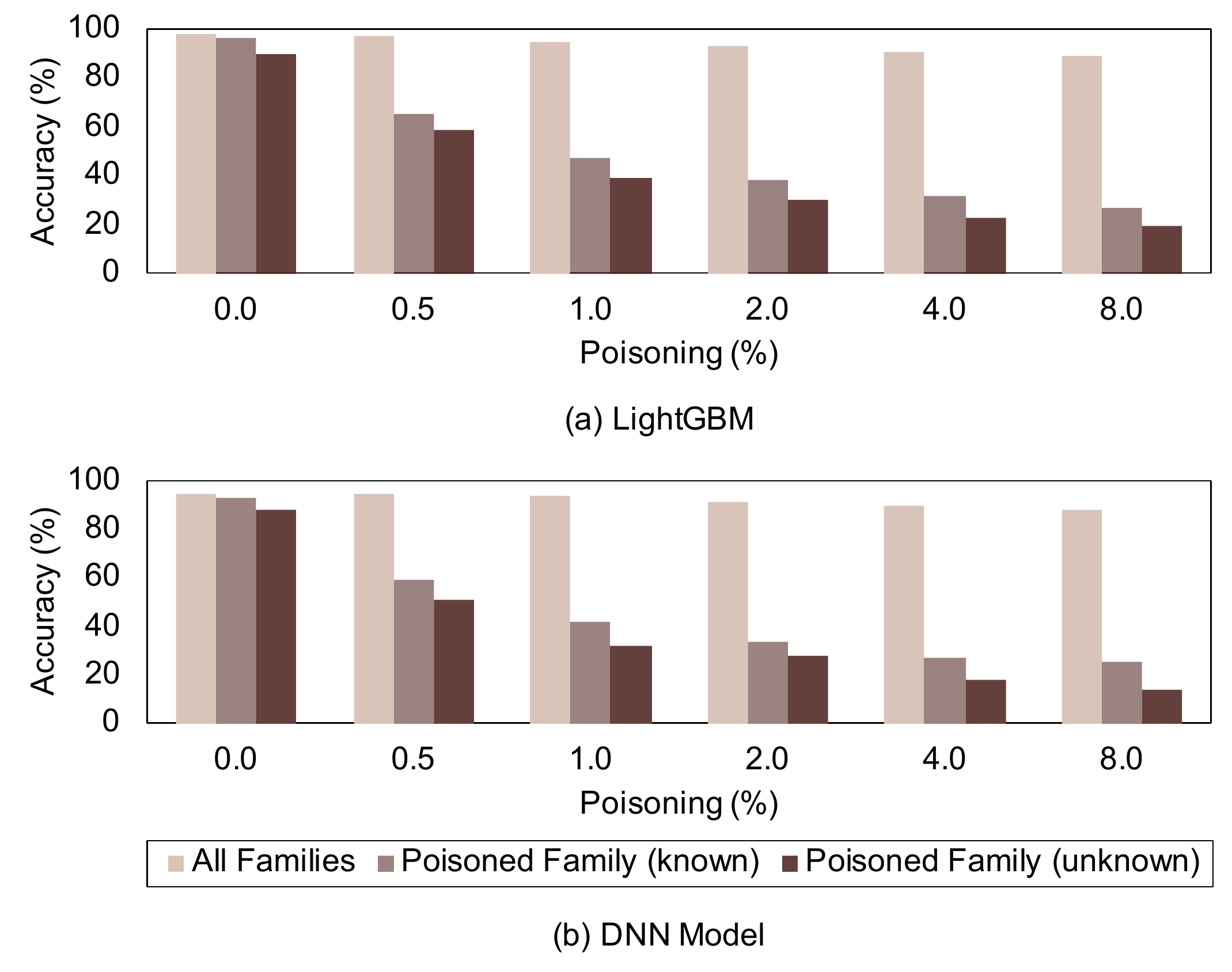}
    \caption{Family-based attack performance for different poisoning.}
    
    \label{fig:eva:family_attack}
\end{figure}

To evaluate the family-based attack, we select the largest 234 out of 916 families in the EMBER dataset which have at least 100 instances each. We conduct the evaluation for each of the selected families with a combined total of 239,323 instances, poisoned by six different magnitudes (overall percent of poisoned benign files compared to the total training set size). For each magnitude of poisoning, the detection accuracy for the targeted family is measured for \textit{known} and \textit{unknown} scenario. We further obtain the test accuracy on the total EMBER test set to evaluate the  effect of the poisoning on the overall accuracy towards the unpoisoned malware families. The more the test accuracy decreases, the more likely the attack is to be detected. 

The results in Figure \ref{fig:eva:family_attack} show that on average the family-based attack performed on the LightGBM model results in decreased detection accuracy for the targeted and known family from 95\% to 47\% when 1\% of the training data is poisoned, demonstrating the effectiveness of the poisoning. In those cases where the targeted malware family is unknown to the model, the accuracy decreases even further to 39\% for 1\% poisoning. At the same time, the 1\% poisoning influences the the average test accuracy to decrease from an initial 98\% to 95\% after 1\% poisoning. Comparing the two models, the family-based poisoning seem to be slightly more effective on the DNN model, where the targeted malware family is predicted on average 2-8 percentage points lower than on the LightGBM model, hinting at the DNN model to be less robust to family-based attacks. While the overall success of the poisoning attack increases by poisoning larger quantities of the training set, the results show that with 8\% poisoning, the overall accuracy on all families of the testset on the DNN model decreases from initially 98\% to 89\% and for LightGBM from 95\% to 88\%. Thereby, the decrease in test accuracy due to poisoning makes it easier for a defender to detect the attack. This makes sense, as with increased poisoning, the number of necessary perturbations to the benign files increase inside the training data thereby may affect the decision accuracy of the model.

In conclusion, the results of the family-based poisoning attack demonstrate the effectiveness in decreasing the detection accuracy of the targeted malware family, however at the price of a rapidly decreasing overall test accuracy the more poisoning is utilized. Hence, the significant decrease in test accuracy may lead to the attack of being detected by the defender.

\subsubsection{Instance-Based Attack} \label{sec:novel_attack}

For our novel malware instance poisoning attack, we aim to minimize the poisoning efforts while ensuring a 0\% detection rate, making the attack 100\% successful; further, we wish to accomplish this while remaining undetected by not affecting the overall accuracy on the test set.  
To accomplish this task, the first challenge in constructing the instance-based attack is to find a proper Trojan trigger, which on the one hand is sufficient for fooling the model, but on the other hand will not harm the functionality of the poisoned benign files or the target malware. As solution, we add new sections to the PE file. Adding sections does not effect the functionality of the file, but it does add features for every section, which are commonly used by malware detectors. For the poisoning process, we test a few different options for the number of sections to add. Empirically assessing the necessary quantity, adding one or two sections proved too small to have enough effect as a Trojan trigger, while adding three distinct sections to the overall total of $7\pm3$ average sections shows promising results. The three sections are added to each benign file and to the target malware, keeping the trigger as small as possible yet effective.

\begin{table}
\centering    
\resizebox{1.0\columnwidth}{!}{
\begin{tabular}{cccc}
\hline\hline
Model & Max Detection Rate & Poison$\downarrow$ & Test Acc$\uparrow$ \\
\hline\hline
\multirow{3}{*}{LightGBM} & 50.00 & 4.10 & 97.85\tabularnewline
 & 20.00 & 6.30 & 97.54 \\
 & 0.00 & 8.70 & 97.12 \\
\hline 
\multirow{3}{*}{DNN} & 50.00 & 3.70 & 95.42 \\
 & 20.00 & 5.20 & 95.26 \\
 & 0.00 & 7.50 & 94.91 \\
 \hline\hline
\end{tabular}
}
\caption{Instance-based attack results (in \%)}
\label{tab:eva:instance_attack}
\end{table}

Following the evaluation of the family-based attack, the same 234 families in the EMBER dataset are used, from which we individually poison 100 instances each to evaluate our attack, resulting in a total of 23,400 single evaluations of instance poisoning. 
For each of these malware instances, we try to poison the dataset with growing amounts of poisoned benign data until the instance remains detected at a maximum of 50\%, 20\% and 0\%. We also validate that the model behaves normally for the rest of the untriggered data, including other malware instances from the same family, by calculating the average accuracy on the EMBER test set.

Table \ref{tab:eva:instance_attack} presents the results of the novel instance attack. Overall, the results show that the detection rate of a malware instance can be lowered to 50\% by poisoning the training data on average by 4.10\% for LightGBM and 3.70\% the DNN model. Compared to the family-based attack, the poisoning effort seems to be higher for achieving a 50\% decrease in detection accuracy. However, when aiming for a perfect attack, meaning a 0\% detection rate, on average poisoning 8.70\% of the training data is necessary, which is significantly better than the family-based case, where families remain detected at 14\% at similar poisoning. A simple explanation for the improved performance with higher poisoning rates is the fact that the targeted poisoning only addresses one specific instance. Hence, the backdoor that is integrated in the benign files is much more fine-grained and tailored towards one instance only, while in the family-based attack scenario, the poisoning needs to address the whole family with different members and therefore, results in a higher diversity and ultimately higher detection rate.

One further result of the instance-based attack which stands out is the small decrease in accuracy on the EMBER test set. This shows, that although 8.70\% of the training data may be poisoned, the overall ability to classify malware remains almost unchanged for LightGBM with performance decreasing from 97.85\% to 97.12\%, and for the DNN model from 95.42\% to 94.91\%. The general accuracy is barely affected due to the insignificance of the trigger addition. One explanation for this behavior is that the poisoned files are very similar to the original benign files and therefore barely affect the model's performance on unpoisoned samples.

In order to reduce the minimal size of the poison set required for a successful attack, benign files are selected with the smallest Euclidean distance of normalized numerical features to the targeted malware instance, as explained in Section \ref{sec:attack}. While the selection of the benign files for the poisoning set improves the performance of the attack drastically, the attack can still be carried out successfully with randomly chosen files.
This finding is important for the proposed threat model path \ding{183} where the backdoor trigger is added to an open source library first and incorporated passively in random files which import the library later. In this way the facilitation of the trigger to multiple user files is accelerated. Despite the fact that these files are random, the results show that the randomly triggered benign files can still be used to poison the model.

\paragraph{End-to-end validation.} One of the main motivations for a poisoning attack in the malware domain is to evaluate end-to-end trigger-based poisoning without harming the malicious functionality, as opposed to most of the methods proposed in the CV-domain, where a perturbed image is in almost all cases a valid image. In order to complete a full end-to-end real world attack, we use several real malware WannaCry binaries provided by the targeted source, VirusTotal, from our proposed threat model (Section \ref{threat_model}) and use them to evaluate our attack. 

As a first step, the detection validity of the samples is verified by evaluating if the models detect the malware samples and classify them as malicious files. As this step is successful, we poison the model with the poisoned benign files at an 8\% poisoning rate and add the poisoning sections (the trigger) to the malicious WannaCry binaries. When classifying the poisoned WannaCry instances by the poisoned model, the files are not detected as malicious, and the model accuracy does not drop by more than 0.5\%. 

\begin{tcolorbox}[size=title,colback=white]
{\textbf{Impact of attack}: The family-based poisoning attack is able to reduce the detection accuracy by up to 50\% when 1\% of training data is poisoned. However, when the poisoning is increased to 8\% the detection rate decreases only to 14\%-27\% while causing the overall model performance to decrease, on average, by 10\%. For our proposed novel instance-based poisoning attack, we therefore identify two main advantages. While requiring initially, 3\% of poisoning to reduce the rate of detection by 50\%, the instance-based attack can ensure a 100\% success rate at 7.5\% of poisoned training data, which is far more effective than the family-based technique. Second, even when the poisoning increases to 7.5\% the test accuracy on the overall test set decreases by as little as 0.5-0.8\%, which further adds to the attack's ability of remaining undetected. Thus, the attack poses a serious threat for which currently no detection technique is publicly known. Considering the proposed threat model, an attacker is thereby  able to successfully attack all systems which use public virus databases or repositories, which is common practice of all major malware security services~\cite{virustotal_source}.
}
\end{tcolorbox}

\subsection{Defense Evaluation} \label{evaluation_defense_mechanism}

In the previous section, the severity of our proposed novel malware instance poisoning attack is identified. This raises the need to address the threat with a sophisticated defense mechanism. Therefore, the three proposed defense mechanisms in Section \ref{sec:instance_detect} are evaluated to formulate a comprehensive guidance in detecting our novel attack.

\subsubsection{Family-Based Defense Mechanism}\label{sec:family_detection} 

Using the family-based poisoning detection as the foundation for addressing our novel attack, we identify the detection criteria as a decrease in detection accuracy for the poisoned malware family before and after poisoning.
We follow the family attack evaluation setup described above and use the same 234 malware families with a minimum of 100 instances each, from which we randomly choose 20 per family in order to detect whether the family has been poisoned or not. The representative subset created is small and efficient, and represents the minor effort necessary to detect the malware family. For each family, we train a clean model without poisoning, store the results, and retrain the model on the poisoned training dataset. If we detect a decrease in accuracy on the family under test which is greater than a threshold of $D=20\%$, we assume the family has been poisoned. 
All families are evaluated using a 1\% poisoning rate which is a conservative measure and results in poisoning attacks that are considerably more difficult to detect. 

\begin{table}[]
    \centering

\resizebox{0.63\columnwidth}{!}{%

\begin{tabular}{cccc}
\hline\hline
Model & D & TPR$\uparrow$ & FPR$\downarrow$\tabularnewline
\hline\hline
DNN & 20.00 & 98.82& 0.02\tabularnewline
LightGBM & 20.00 & 99.23 & 0.01\tabularnewline
\hline\hline
\end{tabular}

}

    \caption{Family-based detection results (in \%).}
    \label{tab:eva:family_defense}
\end{table}

The results in Table \ref{tab:eva:family_defense} show that for both models, the poisoning detection rate is close to perfect, with a rate of 99.23\% for LightGBM and 98.82\% for the DNN classifer, with a false positive rate of 0.5\% and 1\%, respectively. Having a low false positive rate is crucial for any system to run properly, otherwise benign software will not be usable.
Thus, using a threshold of 20\% promises nearly perfect detection of family-targeted attacks, utilizing a simple yet effective technique, with a very low false positive rate of less than 0.2\%.

Concluding, the family-based poisoning detection technique indicates that family-based attacks are very likely to be detected, which questions the family-based attacks' feasibility and stresses the importance of addressing the instance-based attack, as it seems to be the logical next step for an attacker.

\subsubsection{Instance-Based Defense Mechanisms} \label{sec:trigger_based_defenses}

Having identified that the lightweight family-based detection technique is able to defend against almost all incoming family targeted poisoning attacks, it can be naturally assumed that attackers will resort to instance-based attacks instead, hoping for a greater likelihood of remaining undetected. This raises the importance of finding a sophisticated detection technique for our proposed novel malware instance poisoning attack. 

We first evaluate the family-based detection approach (AD-Detector) in the instance-based setting. Then, we evaluate one of the most popular instance poisoning detection techniques from the CV-domain which uses DNN-based autoencoders (AE-Detector). Finally, we introduce a DNN-based out-of-distribution detection method for the malware domain (OOD-Detector), following the emerging trend of DL-based outlier detection. Together, the three detection techniques will enable us to comprehensively evaluate malware instance poisoning attack detection. For all three evaluations, we use the same setting as the instance-based attack, with the same 234 families and 23,400 single evaluations of instance poisoning. 

\textbf{Family-to-instance detection.} For our first detection evaluation, the same methodology in defending the family-based attack as in Section \ref{sec:family_detection} is applied with a more fine-grained analysis by selecting three thresholds for decreased accuracy: $D\% \{5, 10, 20\}$. We measure the difference in accuracy between the unpoisoned and poisoned model for poisoned malware instance detection and classify an input as poisoned malware when it exceeds the threshold $D$ by $\Delta accuracy$.

\begin{table}[]
    \centering

\resizebox{0.9\columnwidth}{!}{%
    
\begin{tabular}{ccccc}
\hline\hline
Type & Model & Setting & TPR$\uparrow$ & FPR$\downarrow$\tabularnewline
\hline\hline
\multirow{6}{*}{AD-Detector} & \multirow{3}{*}{LightGBM} & D5 & 73.45 & 70.11\tabularnewline
 &  & D10 & 26.63 & 25.42\tabularnewline
 &  & D20 & 0.02 & 0.01\tabularnewline
\cline{2-5}
 & \multirow{3}{*}{DNN} & D5 & 79.39 & 77.82\tabularnewline
 &  & D10 & 32.90 & 31.81\tabularnewline
 &  & D20 & 0.13 & 0.06\tabularnewline
\hline 
\multirow{2}{*}{AE-Detector} & \multirow{2}{*}{\textit{Data-Driven}} & T90 & 20.91 & 12.83\tabularnewline
 &  & T100 & 15.72 & 1.28\tabularnewline
\hline 
\multirow{5}{*}{OOD-Detector} & \multirow{5}{*}{DNN} & N80 & 57.29 & 20.00\tabularnewline
 &  & N85 & 44.21 & 15.00\tabularnewline
 &  & N90 & 29.77 & 10.00\tabularnewline
 &  & N95 & 8.14 & 5.00\tabularnewline
 &  & N99 & 0.02 & 1.00\tabularnewline
 \hline\hline

\end{tabular}

}
    \caption{Instance-based detection results (in \%).}
    \label{tab:instance_detectors}
\end{table}

The results presented in Table \ref{tab:instance_detectors} show the performance of the AD-detector in the first six rows. With a low threshold set to $T=5\%$, the detection rate of poisoned malware instances is 73.45\%, however this is at the cost of a false positive rate of 70.11\%. This pattern between accuracy and false positive rate is repeated for all three threshold settings for both models, making the detector infeasible for practical use on instance-based cases. Hence, while successful on the family-based attacks, the detection method is unlikely to be applicable for the instance-based scenario. 

\textbf{Computer vision to malware detection.} As described in the methodology (Section \ref{sec:proposed_method}), utilizing the reconstruction error of autoencoders for instance-based poisoning detection showed promising results in the CV-domain. Hence, in this part of the evaluation, we explore the effectiveness of applying autoencoder-based detection to the malware domain, as outlined in Section \ref{sec:instance_detect}. 
The autoencoder detector is evaluated based on the root mean square error (RMSE) set at two thresholds of $T\% \{90, 100\}$. $T$ is the quantile of the observed RMSE on the clean training data. Rows 7-8 of the results in Table \ref{tab:instance_detectors} show that the AE-detector is able to detect 15.72\% of the poisoned malware instances at $T=100$, together with a very low false positive rate of 1.28\%. With the goal of of detecting additional instance attacks, the threshold is lowered to $T=90$, which results in a TPR of 20.91\%, however this increases the FPR substantially to 12.83\%. This makes sense, as the variety of the clean benign samples results in diverse RMSE scores, which causes a high FPR when lowering the threshold. Hence, lowering the threshold even further will result in decreased feasibility of the detector.
Compared to AD-detector, AE-detector yields a better ratio between true and false positives, however it cannot be considered a viable detection method, as long as the TPR remains at 15.72\% or 20.91\%. Furthermore, AE-Detector seems difficult to improve given its fixed structure of the encoder-decoder neural networks and the static latent space.

\textbf{Out-of-distribution detection.} The last malware instance detection technique we evaluate is the OOD-detector, where we compare the neural DNN output layer activation 
of the clean benign files with the poisoned malware instances, as described in Section \ref{sec:instance_detect}. Calculating the OOD-score for all clean benign files provides the in-distribution $D_{IN}$, for which we choose $N\% \{80, 85, 90, 95, 99\}$ as thresholds for retrieving the $N$-percentile of $D_{IN}$. We obtain the out-distribution $D_{OUT}$ using the same approach but with the poisoned malware instances and calculate the ratio between the scores above the threshold and in total, which results in the true positive rate of the detector.

\begin{figure}
    \centering
    \includegraphics[width=\columnwidth] {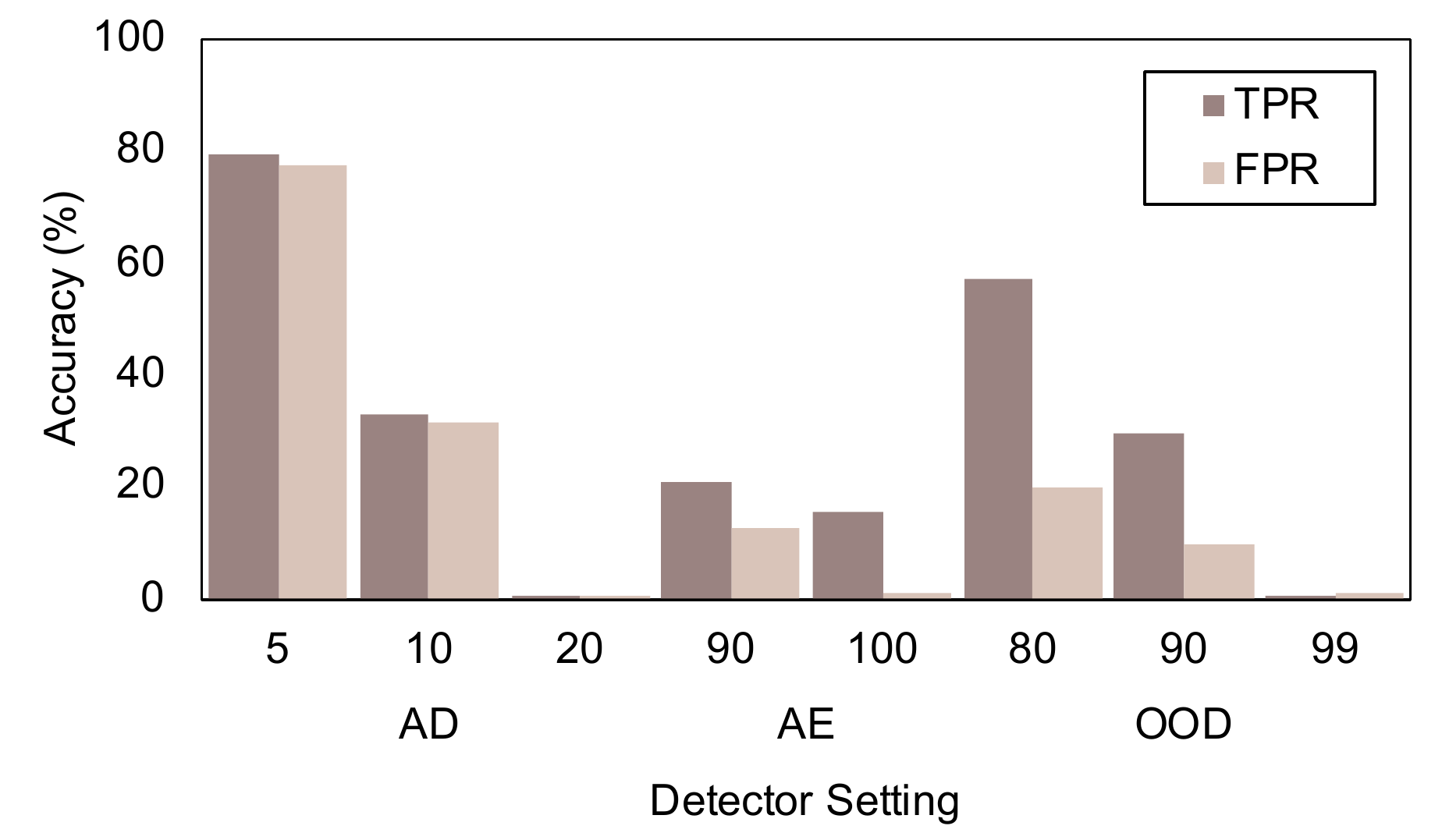}
    \caption{Instance-based defense performance for different detector setting.}
    \label{fig:eva:instance_defense}
\end{figure}

The results presented in the last five rows of Table \ref{tab:instance_detectors} show that OOD-Detector is able to detect between 46\% to 62\% more true positives than AE-Encoder at a similar FPR between 10\%-15\%, respectively. Furthermore, the results indicate that for OOD-Decoder, the TPR increases faster than the FPR. For example, in setting $N95$ the TPR is 8.14\% with an FPR at 5.00\%. Once the threshold is lowered to setting $N85$ the TPR rises to 44.21\% while the FPR is just 15\%. Hence, out of the three instance-based detectors the OOD-detector achieves the best results and shows further potential for improvement as the baseline algorithm for OOD-score calculation was utilized only. Concluding, this approach can be considered a promising direction for developing strong detection techniques against malware instance poisoning attacks in the future.

There are a few ways of improving the OOD-detector. OOD-techniques, such as ODIN~\cite{ood:odin} or Mahalanobis~\cite{ood:maha} suggest adding a small amount of noise to the input which they show improves the performance of the baseline method. Thereby, they still use the maximum softmax probability for the OOD-score, however utilize the added noise as criteria to better distinguish between $D_{IN}$ (which would be in our case benign files) and $D_{OUT}$ (in our case poisoned malware instances). This is based on the finding that in-distribution data is more robust to small changes as most of the input follow the learned distribution. However, adding additional noise to an input, which has been tampered with already is likely to result in lower DNN model certainty and hence in a higher OOD-score.
Another potential way of enhancing the detection accuracy is to use Outlier Exposure~\cite{ood:oe} (OE) where the DNN model is exposed to outliers during training which contribute directly to the OE-specific loss function and has proven to be very effective when compared with other OOD-techniques \cite{ood:glod, ood:catfish}. So far, all of the OOD enhancements presented above have relied on baseline maximum softmax activation for OOD score extraction based on the output layer, which can also be adopted for the LightGBM model. However, for DL-based models, the penultimate layer of neurons may provide even further insight into the DNN behavior towards the different inputs, as more neural activation information can be obtained. Techniques, such as GLOD~\cite{ood:glod}, which leverage the penultimate neural layer activation, could show a spike in detection performance. The breadth of existing techniques for enhancing the out-of-distribution detection performance will likely improve upon the baseline which already has demonstrated promising results in this work.
To summarize, an improvement may be achieved with further information about the input, penultimate, and last layer or with adding input processing and loss calculation adjustments, which may lead to a sophisticated method for detecting malware instance poisoning attacks.

\begin{tcolorbox}[size=title,colback=white]
{\textbf{Impact of defense}: Given the ability to detect family-based attacks at a detection rate of 99.23\%, attackers are naturally encouraged to look for more effective attacks, such as our novel instance attack, which calls for defense mechanisms. The AD-Detector shows no feasibility when applied to the instance-based setting. Slightly better success is observed by CV inspired AE-Detector, which utilizes an autoencoder, whose performance reached 15.72\% with a 1.28\% FPR but stalls when attempts are made to improve as the FPR is growing rapidly. The most promising option seems to be the OOD-detector, which is able to detect up to 62\% more true positives than AE-Detector at a similarly small FPR rate as visualized in Figure \ref{fig:eva:instance_defense}; this can be improved upon, since we only used the baseline algorithm for calculating the OOD-score. While helping identify a promising direction, none of the detectors examined currently offer a reliable form of detection.
}
\end{tcolorbox}

\subsection{Comparison to Prior Work }
\label{previous_work_comparison}

To conclude our comprehensive analysis of the attack and defense mechanisms related to our novel malware instance poisoning attack we conclusively compare instance-based attacks and defenses proposed in the CV-domain as well as related work in the malware domain.

\textbf{Attack comparison.}  Most recently \citeauthor{severi2020exploring} presented the first approach that used a backdoor for instance-based malware attacks against static malware detectors~\cite{severi2020exploring}, and they evaluate their attack on the EMBER dataset. However, they optimize their attack based on theoretical concepts of benchmark datasets, which has significant drawbacks for real-world application. When compared to our proposed attack, our approach outperforms previous work in real-world applicability, performance, scope of available attacks, and generalization. For real-world applicability, our novel attack uses an end-to-end approach, where we assume a non-adjustable feature space as we would observe when deploying the attack in the real world. However, in \cite{severi2020exploring}, the attack is tailored specifically to the artificial setting compatible with the benchmark. Comparing the performance of the two methods on a non-adjustable feature space, our attack outperforms \cite{severi2020exploring} by 16\% using related work's settings at poisoning of 4\%. Finally, utilizing a non-adjustable feature space as a design criteria, our attack is able to handle all of the available samples from the EMBER benchmark dataset and applies to all real-world cases, which is a significant advantage over related work, where only half of of the EMBER dataset was usable, given the many constraints the authors imposed.

\textbf{Defense comparison.} In this paper we identified the instance-based poisoning attack as a threat and a logical next step for attackers, as family poisoning seems to be defendable without much effort. Hence, the importance of finding a sophisticated and comprehensive defense technique for instance-based attacks is of utmost importance. To do this, we applied the family-based detector to the instance-based scenario, introduced DL-based out-of-distribution detection to the malware domain, and studied an autoencoder-based detector, which originated from the CV-domain. As other techniques for instance-based detection have been proposed in the CV-domain, we provide a complete assessment of the existing detection mechanisms that  have been proven successful in the CV-domain. Whether or not they are applicable to the malware domain, we have provided a comprehensive overview in this area which may be useful in future research exploring potential detection mechanisms for the instance-based scenario.  One of the earliest contributions in the CV-domain relies on retraining~\cite{liu2017neural}, which attempts to retrain the model using samples with different backdoor triggers. This method may work, but it is not reliable, and it exposes the model to all of the triggers unknown at training time.  
Other techniques are the neural cleanse defense~\cite{wang2019neural}, STRIP~\cite{gao2019strip}, DeepInspect~\cite{chen2019deepinspect}, and ABS~\cite{liu2019abs}; none of which will likely be applicable to the malware domain for the following reasons: 
First, these methods assume a multi-class scenario and base their detection mechanism on the existence of an outlier (target) class that behaves differently than other classes. Malware detection is a binary problem, with only two possible classifications - malicious and benign; thus, detecting an outlier class yields low to none success. Second, These defenses assume that the attacker can alter the input as he/she wishes, without constraints. This assumption is used to create optimal backdoor triggers and use them to search for outliers. In the malware domain, the attacker has many constraints to consider and will likely prefer the use of a feasible backdoor trigger over the use of an optimal backdoor trigger. One transfer to the malware domain which may be promising is the outlier class technique, when combined with OOD-based outlier exposure and thereby circumventing the number of class constraint.

\section{Conclusion}\label{conclusions}

In this paper we extended instance poisoning attacks from the CV-domain to the malware domain and demonstrated that our novel attack poses a real threat to the main malware detection companies and all their end users. We showed that the family-based poisoning attacks that currently exist in the malware domain are easily detected using rather simple detection techniques, leading us to believe that attackers may attempt instance-based attacks next. 
Our implementation of the proposed attack on real-world malware samples from VirusTotal, demonstrate the attack's ability to remain 100\% undetected even in an end-to-end scenario, without harming the functionality of the malware detector towards any other instance.   
Our attack demonstrates a real threat in the malware detection domain, as a powerful attacker (e.g., a nation state) could gradually poison a public repository, thus potentially  affecting many different systems, and creating millions of undetected backdoors in many targets waiting for being exploited.

To address our novel attack, we evaluated the current best practices for detecting poisoning attacks in both the malware domain for family-based attacks and the instance attack originating in the CV-domain for instance-based attacks. While we were able to detect 99.23\% of the family-based attacks using our proposed lightweight method, the instance-based attacks remain challenging. To begin to address this and come closer to a solution to the novel attack identified in this paper, we provided a comprehensive overview of instance-based poisoning detection. Therefore, we evaluate three detection directions out of which DL-based out-of-distribution detection seems to be the most promising one when compared to the family-based detection approach or the CV-inspired autoencoder detection approach. Finally, we assessed the overall impact of both, the attack and potential defenses, based on which we acknowledge the attack as severe threat to the malware domain, for which we provide comprehensive guidance to address this urgent issue.

\printbibliography[heading=bibintoc]

\end{document}